\documentclass[11pt,a4paper]{article}
\pdfoutput=1
\usepackage{jheppubnohead}
\usepackage[utf8]{inputenc}
\usepackage{amsmath}
\usepackage{epsfig}
\usepackage{graphicx}
\usepackage{amssymb}
\usepackage{tabularx}
\usepackage[normalem]{ulem} 
\usepackage{booktabs} 
\usepackage{subfigure}

\providecommand{\proarrow}[0]{\rightarrow}

\providecommand{\dif}[0]{\mathrm{d}}

\providecommand{\proname}[2]{#1 \proarrow #2}
\providecommand{\lrproname}[2]{#1 \leftrightarrow #2}

\providecommand{\abs}[1]{\left\lvert #1 \right\rvert}

\providecommand{\abss}[1]{\left\lvert #1 \right\rvert^2}

\providecommand{\miim}[1]{{\rm Im} \left[ #1 \right]}

\providecommand{\order}[1]{{\cal O} \left( #1 \right)}

\providecommand{\tsph}[0]{T_{sfo}}

\providecommand{\ghor}[2]{\gamma \left(\proname{#1}{#2}\right)}

\providecommand{\ydyeq}[1]{\frac{Y_{#1}}{Y^{eq}_{#1}}}
\providecommand{\ydyeqs}[1]{\frac{Y_{#1}^2}{Y^{eq\, 2}_{#1}}}

\providecommand{\np}[1]{\nu_{#1}^+}
\providecommand{\nm}[1]{\nu_{#1}^-}
\providecommand{\npj}[0]{\np{j}}
\providecommand{\nmj}[0]{\nm{j}}
\providecommand{\innerrightleftarrow}[0]{\rightarrow \mspace{-10.0mu} - \mspace{-10.0mu} \leftarrow}
\providecommand{\innerleftrightarrow}[0]{- \mspace{-10.0mu} \leftrightarrow \mspace{-10.0mu} -}

\providecommand{\innerlongleftarrow}[0]{- \mspace{-10.0mu} \longleftarrow}

\newcommand{\be}{\begin{equation}}
\newcommand{\ee}{\end{equation}}
\newcommand{\bea}{\begin{eqnarray}}
\newcommand{\eea}{\end{eqnarray}}

\title{Helicitogenesis: WIMPy baryogenesis with sterile neutrinos and other realizations}
\author[]{J.~Racker,}
\author[]{N.~Rius}

\affiliation[]{Departamento\ de F\'{\i}sica Te\'orica, Instituto de F\'isica corpuscular (IFIC), Universidad de
Valencia-CSIC \\ 
Edificio de Institutos de Paterna, Apt. 22085, 46071 Valencia,
Spain
}
\emailAdd{racker@ific.uv.es}
\emailAdd{nuria@ific.uv.es}

\keywords{Baryogenesis, Dark matter, Neutrino masses, Beyond Standard Model}
\abstract{We propose a mechanism for baryogenesis from particle decays or annihilations that can work at the TeV scale. Some heavy particles annihilate or decay into a heavy sterile neutrino $N$ (with $M \gtrsim 0.5$~TeV) and a ``light'' one $\nu$ (with $m \ll 100$~GeV), generating an asymmetry among the two helicity degrees of freedom of $\nu$. This asymmetry is partially transferred to Standard Model leptons via fast Yukawa interactions and reprocessed into a baryon asymmetry by the electroweak sphalerons. We illustrate this mechanism in a WIMPy baryogenesis  model where the helicity asymmetry is generated in the annihilation of dark matter. This model connects the baryon asymmetry, dark matter, and neutrino masses. Moreover it also 
complements previous studies on general requirements for baryogenesis from dark matter annihilation. Finally we discuss other possible realizations of this helicitogenesis mechanism.}
\begin{document}
\hfill {\tt IFIC/14-42, FTUV-14-0419}

\maketitle
\section{Introduction}

The nature of the Dark Matter (DM) and the origin of the Baryon Asymmetry of the Universe (BAU) are unknown and both require physics  beyond the Standard Model (SM) to be explained. Also puzzling -or maybe a hint?- is the fact that the energy densities are comparable,  $\Omega_{DM} \sim 5 \, \Omega_B$~\cite{planck13}. The conventional explanations are  unrelated, and 
often involve very different scales of new physics. 
The most popular candidates for DM are Weakly Interacting 
Massive Particles (WIMPs), which provide the so-called ``WIMP miracle": the thermal relic abundance of a stable WIMP is naturally of the order of the observed $\Omega_{DM}$.
Regarding the baryon abundance, 
 the Sakharov conditions to generate dynamically the BAU can be fulfilled  in a variety of extensions of the SM at very different energy scales, ranging from 
 below the electroweak  to the Planck scale. 
  However the similarity of the DM and baryonic energy densities suggests a common origin, 
and such possibility has been extensively studied in recent years.

Most models relating the dark and baryonic matter abundances involve  
Asymmetric Dark Matter (ADM), i.e., the DM we observe today is due to a  particle-antiparticle asymmetry in the dark sector which is somehow tied to the one in baryons
(for extensive reviews see~\cite{petraki13, zurek13}).  
However in the ADM scenario the ``WIMP miracle" is lost. This has motivated several attempts 
to find a mechanism that preserves the natural DM  relic density of a WIMP and 
at the same time relates the dark and baryonic matter abundances~\cite{McDonald:2011zza, McDonald:2011sv}.
One possibility is the existence  of several WIMPs, at least one of them stable which will make up the DM of the Universe and 
other(s) long-lived, which will generate the BAU in their out of equilibrium decay~\cite{Davidson:2012fn, cui12}. 
In a more minimalistic mechanism, dubbed WIMPy Baryogenesis (WB),   the BAU is generated directly in the annihilation of a stable WIMP~\cite{cui11}.
The  phenomenology of several WB models has been studied in~\cite{bernal12} and 
conditions for generating the observed $\Omega_{B}$ and $\Omega_{DM}$ via this mechanism
analyzed in detail in~\cite{bernal13}.

One of the challenges of WB is that the BAU must be produced at temperatures $T \sim m_\chi$ -with $m_\chi$ the DM mass- which are very low for thermal baryogenesis mechanisms (usually $m_\chi \lesssim {\rm few \; TeV}$ and in any case $m_\chi < 340$~TeV~\cite{griest89}). At such temperatures the processes responsible for the CP even phase in the reaction generating the BAU are typically very fast compared to the Hubble rate, and therefore they erase most of the cosmic asymmetry (for recent detailed discussions on this subject see~\cite{bernal13, racker13}). One way out of this problem is to include massive particles so that the washout processes decouple exponentially. To the best of our knowledge this idea was first fully explored in~\cite{cui11} for baryogenesis from DM annihilations, and  
studied in~\cite{racker13} for baryogenesis from particle decays. In these works the annihilations -or decays- directly produced a baryon or lepton asymmetry in SM fields, and given that the mediators of the annihilations -or the decaying particles- were singlets, the massive fields had some SM charges and could also store an asymmetry. In turn this asymmetry had to vanish exponentially without canceling the SM baryon -or lepton- asymmetry, which lead to some complications in the implementation of the mechanism (as the need for a light sterile dark sector or a very fast interaction violating some of the charges of the massive field). 

In this work we propose a variation of the above scenario: the massive particles responsible for the exponential suppression of the washout are Majorana fermions and therefore they do not have any conserved charge, avoiding the complications just mentioned. 
 It can be realized in both, baryogenesis from DM freeze out and from heavy particle out-of-equilibrium decay. 
We 
illustrate this mechanism in a WB model where the DM annihilates into sterile neutrinos, which in turn are responsible for neutrino masses via the type I seesaw. In this way we address the possibility of relating the DM and BAU problems with yet another puzzle that requires physics beyond the SM: neutrino masses. Furthermore our work complements previous studies~\cite{bernal13} on general conditions for having baryogenesis from DM annihilation.

The basic requirement for generating the BAU is the existence of heavy   sterile neutrinos $N$   ($M \gtrsim 0.5$~TeV) and 
also  lighter ones $\nu$ (with $m \ll 100$~GeV), both of them interacting 
with the DM via SM singlet scalars.   
WIMP annihilations to the sterile neutrinos violate CP, 
 generating an asymmetry among the two helicity degrees of freedom of $\nu$. 
 This asymmetry is transferred to SM leptons via fast Yukawa 
interactions, which should be in equilibrium prior to the electroweak phase transition 
to ensure that a baryon asymmetry is also induced by the sphaleron 
$(B+L)$-violating interactions. Moreover, since no asymmetry accumulates in the heavy sector 
$N$, some of the requisites of the original WB models -a $Z_4$ symmetry and a light sterile dark sector- are automatically avoided.

 Given that the mass scale of the sterile neutrinos is unconstrained and its origin unknown, it is worthwhile to explore the consequences of having several mass scales in the sterile sector without any theoretical prejudice.  Therefore we first 
 adopt a purely phenomenological perspective, 
 illustrating the proposed mechanism by means of a  minimal 
model.

It is however tempting to justify the hierarchical sterile neutrino mass spectrum by some 
broken, global or local, symmetry. Moreover, there are neutrino mass models  in which this is actually the case, as in the double  seesaw scenario~\cite{mohapatra86}.
We then discuss a  realization of the 
 ``helicitogenesis" mechanism within the double seesaw framework, 
 in a model with spontaneously broken $U(1)_L$ symmetry 
 ($U(1)_{B-L}$ in the gauged case), in which the DM  is charged under lepton number.

Notice that a similar helicity asymmetry in the  SM-singlet Majorana neutrinos, also 
transferred to the SM lepton sector by  fast Yukawa interactions,  is the basis of 
baryogenesis via neutrino oscillations originally proposed in~\cite{akhmedov98, asaka05}. In such 
case the helicity asymmetry is due to the CP-violating coherent neutrino oscillations, while 
in our scenario it is generated in the CP-violating DM annihilation or CP-violating decay
of a heavy particle. Therefore the requirements  $m \ll 100$~GeV and some sterile-active neutrino
Yukawa interactions in equilibrium before the electroweak phase transition are 
common to all mechanisms, while we have extra sources of CP violation.

The paper is organized as follows.   
In section~\ref{sec:model}  we describe the basics of the helicitogenesis mechanism and implement it in a minimal WB model.
In section~\ref{sec:helicity} we write   
the set of Boltzmann Equations (BEs) relevant for the generation and evolution of the 
helicity asymmetry in the sterile neutrinos.  
Section~\ref{sec:other}  is devoted to 
other realizations of 
helicitogenesis in the context of a neutrino mass model with spontaneous 
$U(1)_L$ symmetry breaking,  
and we conclude in section~\ref{sec:conclusions}.

\section{The mechanism and a WIMPy leptogenesis model}
\label{sec:model}
In WB~\cite{cui11} the DM, $\chi$, is a weakly interacting massive particle whose relic density is determined by the freeze out of some annihilation process $\proname{\chi \chi}{\bar \Psi f}$, with $f$ a SM fermion and $\Psi$ a heavy exotic particle. The amplitude for the process $\proname{\chi \chi}{\bar \Psi f}$ contains  a CP odd phase coming from complex couplings and a CP even phase from the absorptive part of one loop contributions, therefore it violates CP. Moreover, depending on whether $f$ is a SM quark or a lepton, the interaction $\proname{\chi \chi}{\bar \Psi f}$ violates SM B or L, respectively. In this way all Sakharov conditions are satisfied and some baryon or lepton asymmetry is produced in the annihilation of DM. If $\Psi$ is heavy enough, $m_\Psi \gtrsim m_\chi$~\cite{cui11}, the processes that can potentially washout the asymmetry -most notably $\lrproname{\bar \Psi f}{\Psi \bar f}$- are Boltzmann suppressed, hence a significant amount of matter asymmetry may survive. 

However, the annihilation $\proname{\chi \chi}{\bar \Psi f}$ also generates an asymmetry in the $\Psi$ sector and it is not trivial to avoid a cancellation of the total matter asymmetry after $\Psi$ disappears from the thermal bath. This seems to force into building more complicated models. E.g. in the original work~\cite{cui11} the $\Psi$ decay into a light hidden sector, while decays into SM particles are forbidden by a $Z_4$ symmetry. Then in~\cite{bernal13} it was shown that WB could work without a light hidden sector or a discrete $Z_4$ symmetry, but still the problem associated to the asymmetry in $\Psi$ had to be solved complicating the models in some other ways. Here we present a WIMPy model where the role of $\Psi$ is played by heavy Majorana fermions (subsequently called $N_i$) which lack a conserved charge to store asymmetry~\footnote{A massive real scalar could also be used to suppress the dangerous washouts in a similar way, but a different baryogenesis model would be necessary to implement this option.}. Therefore no additional fields beyond those participating in the annihilation of DM must be added. In addition, this model yields a close connection to light neutrino masses.

The SM is extended with some singlet real scalars, $S_a$, and Majorana fermions, $\chi, N_i, \nu_j$, with $\{a,i,j=1,\dots\}$, together with a discrete $Z_2$ symmetry to ensure the stability of the DM. The DM candidate $\chi$ is the only odd particle under $Z_2$. It could also be a Dirac singlet, but we choose it Majorana to minimize the number of new degrees of freedom. 
The $N_{i}$ and  $\nu_j$ are sterile neutrinos with ``high'' ($M_i \sim {\cal O}$(TeV)) and ``low'' 
($m_j \ll 100$~GeV) masses, respectively
\footnote{Although in this model the species $\nu_j$ and $N_i$ differ only in their masses, we denote them by different symbols to emphasize their distinct roles for leptogenesis.}. 
 In the basis which yields a diagonal Majorana mass matrix with real and positive entries, the most general renormalizable Lagrangian with the given fields and symmetries reads
\begin{equation}
\label{eq:lag}
\begin{split}
-L = &- L_{SM} - L_{kin} + V(S_a,H) \\
&+ \frac{1}{2} \Big\{ m_\chi \bar\chi \chi + M_i \bar N_i N_i  + m_j \bar \nu_j \nu_j  \Big\} \\
& + \frac{1}{2} \Big\{ \lambda_{\chi a} S_a \bar \chi P_R \chi + \lambda_{N a i j} S_a \bar N_i P_R N_j + \lambda_{\nu a i j} S_a \bar \nu_i P_R \nu_j 
+ {\rm h.c.} \Big\} \\
  & + 
  \Big \{  
   \lambda_{a i j} S_a \bar N_i P_R \nu_j + 
  h_{N \alpha i} \tilde H \bar \ell_\alpha P_R N_i + h_{\nu \alpha j} \tilde H \bar \ell_\alpha P_R \nu_j + {\rm h.c.} 
  \Big\}
   \; ,
\end{split}
\end{equation}
where there is an implicit sum over repeated family indices, $\ell_\alpha$ are the leptonic $SU(2)$ doublets, $H$ is the Higgs field ($\widetilde H_2 =i\tau_2 H_2^*$, with $\tau_2$ Pauli's second matrix), $V(S_a,H)$ is the scalar potential and $P_{R,L} =  (1 \pm \gamma_5)/2$ are the chirality projectors. Latin indices denote sterile neutrinos while Greek indices refer to the SM lepton doublets.
All Majorana fields $\xi$ ($\xi=\chi, N_i, \nu_j$) satisfy $\xi^c=\lambda_\xi \xi$, with $\lambda_\xi$ a phase factor. Notice that the Yukawa matrices 
$\lambda_{N a}, \lambda_{\nu a}$ are symmetric. 

The key for having baryogenesis in this model is that $m_j \ll \tsph$ at least for one species 
$\nu_j$, where $\tsph = {\cal O}$(100~GeV) is the temperature at which the electroweak sphalerons freeze out. Then an asymmetry among the two helicity degrees of freedom of $\nu_j$, $\npj$ and $\nmj$, can be generated from the annihilation of DM $\proname{\chi \chi}{N_i \nu_j}$, in the same way as the lepton or baryon asymmetry is created from $\proname{\chi \chi}{\bar \Psi f}$ in previous WB models~\cite{cui11, bernal12, bernal13}. As long as some of the Yukawa couplings of $\nu_j$, $h_{\nu \alpha j}$, are large enough, the helicity asymmetry in the $\nu_j$ is efficiently transferred to the SM lepton sector. In turn, this is partially transformed into a baryon asymmetry by the sphaleron processes. Once these decouple at $\tsph$, the BAU is frozen. 
Notice that if $m_j \neq 0$, the helicity depends  on the reference frame. We will be always 
working  in the thermal bath rest frame.

This baryogenesis scenario requires at least one species of $N's$ and $\nu's$,
and two real scalars to have CP violation, $S_1$ and $S_2$ (actually we will explain later that there can be a CP odd phase with just one scalar, contrary to previous WB models, but the amount of CP violation is most likely too small to have successful baryogenesis). Next we indicate the approximate range of values that the parameters in the Lagrangian~(\ref{eq:lag}) can take for this baryogenesis mechanism to be successful.

\begin{itemize}
\item $m_{\chi}$: To generate the asymmetry before sphalerons freeze out, the DM has to start annihilating well above $\tsph$, hence $m_{\chi} \gtrsim 1$~TeV.
\item $m_{S a}$: Although the asymmetry could also be produced in the decays of $S_a$ (see~\cite{cui11}), we are interested  in the case that the asymmetry is mainly produced in the annihilation of DM, hence we impose  that the masses of the singlet scalars, $m_{S a}$, are
$m_{S a} \gtrsim m_{\chi}$ (in this way the CP conserving annihilation channel $\proname{\chi \chi}{S_a S_a}$ is negligible).
\item $M_i$: The heavy $N_i$ are introduced to have a Boltzmann suppression $e^{-M_i/T}$ of washouts that can be very fast when baryogenesis occurs at low temperatures (see~\cite{cui11} and~\cite{bernal13,racker13} for detailed discussions on this point). For this Boltzmann suppression to be efficient $M_i \gtrsim (0.5 - 1) \, m_\chi$. In addition, $M_i < 2 m_\chi$ to allow for DM annihilations when $\chi$ becomes non-relativistic.     
\item $m_j$: To create an helicity asymmetry in the $\nu$-sector it is necessary that  $m_j \ll \tsph$. In Sec.~\ref{sec:helicity} this issue will be studied in more detail.
\item $\lambda_{\chi a}, \; \lambda_{a i j}$: These couplings must be ${\cal O}$(1) for having enough CP violation and a correct DM relic abundance. More precisely, it is the imaginary part of $\lambda_{\chi a}$ that has to be large, so that there is a sufficiently fast, not velocity-suppressed annihilation rate.
\item $\lambda_{\nu a i j}$: They induce washouts of the helicity asymmetry that are not Boltzmann-suppressed, e.g. via the reaction $\lrproname{\npj \npj}{\nmj \nmj}$. For these processes to be slow enough $\displaystyle \abs{\lambda_{\nu a i j}} m_\chi/m_{S a}$ $\ll 10^{-3}$.
\item $\lambda_{N a i j}$: They do not play an important role because the corresponding processes are Boltzmann suppressed, hence they are unconstrained.
\item $h_{\nu \alpha j}$: It is crucial that there be at least one fast Yukawa interaction between the $\nu_j$ and $\ell_\alpha$, and therefore at least one coupling $h_{\nu \alpha j} \gtrsim 2 \times 10^{-7}$~\cite{campbell92,cline93}.
\item $h_{N \alpha i}$: They mediate washout processes like $\lrproname{\ell_\alpha H}{\bar \ell_\alpha \bar H}$, 
which should be slow at $T \gtrsim \tsph$. As for the $\lambda_{\nu a i j}$ couplings, this requirement is satisfied for $\abs{h_{N \alpha i}} m_\chi/M_i \ll 10^{-3}$.
\end{itemize}

In addition  the heavy singlet sector must be populated at $T \gtrsim m_{\chi}$. This can be achieved by some fast interaction connecting the sterile and SM sectors, like one among the $S_a$ and the Higgs or the Yukawa interactions 
between the $N_i$ and $\ell_\alpha$.

The allowed region in the parameter space of $m_{\chi}, m_{S a}, M_i, \lambda_{\chi a}$ and $\lambda_{a i j}$ is very similar to previous models of WB~\cite{cui11,bernal12,bernal13} (with $N_i$ playing the role of the heavy exotic annihilation product); very roughly it consists of masses above $\sim$ 1~TeV and ${\cal O}(1)$ couplings. 
On the other hand, if helicitogenesis occurs in the decay of the lightest scalar, $S_1$, the conditions for 
generating the $\nu$ helicity asymmetry 
are analogous to those for standard leptogenesis 
at the TeV scale, basically   
$\lambda_{1 i j} \lesssim 10^{-7}, m_{S a} \gtrsim 1$~TeV, and $M_i \gtrsim 0.5$~TeV~\cite{racker13}.
Hence we are not going to develop these issues further. 
Instead we will concentrate on the constraints imposed by leptogenesis via helicitogenesis and the connection with light neutrino masses: 

\begin{itemize}
\item[(i)]

The $N$'s and $\nu$'s must decay before Big Bang Nucleosynthesis (BBN)
 to avoid observational constraints. 
 
This requirement is not difficult to accomplish for the heavy neutrinos $N_i$.
One possibility is that $h_{N \alpha i}$ be non-negligible to allow for the decay $\proname{N_i}{\ell_\alpha H}$, but at the same time small enough for the washout processes like $\lrproname{\ell_\alpha H}{\bar \ell_\alpha \bar H}$ to be slow at $T \gtrsim \tsph$. This condition is easy to satisfy given that the rate of this last process is $\propto h_{N \alpha i}^4$, while the rate of the former is $\propto h_{N \alpha i}^2$.
Another possibility could be to choose $\lambda_{\nu a i j}$ large enough to induce three body decays like  $\proname{N_i}{\nu_j \nu_j \nu_j}$, but not as large as to have fast washouts $\lrproname{\npj \npj}{\nmj \nmj}$ (again note that the rate of $\proname{\npj \npj}{\nmj \nmj}$ is $\propto \lambda_{\nu a j j}^4$ while the rate of $\proname{N_i}{\nu_j \nu_j \nu_j}$ is $\propto \lambda_{\nu a j j}^2$).

The  main decay modes  of the (lightest) $\nu_j$ are 
$\nu_j \rightarrow \nu_\alpha \nu_\beta \bar {\nu}_\beta,
\nu_j \rightarrow \nu_\alpha e^-_\beta e^+_\beta, 
\nu_j \rightarrow \nu_\alpha q_\beta  \bar {q}_\beta$, 
via $Z$ exchange, $\nu_j \rightarrow e^-_\alpha e^+_\beta \nu_\beta, 
\nu_j \rightarrow e^-_\alpha q_\beta \bar {q}'_\beta$ via $W$ exchange, 
and the corresponding CP-conjugate processes, 
where $e_\alpha$ denotes the charged leptons $e,\mu,\tau$  
 and $q_\beta$ stands for the SM quarks, except the top. The $\nu_j$  decay width 
is given by 
\begin{equation}
\label{eq:nudecay}
\Gamma_{j} = \frac{G_F^2 m_j^3}{192 \pi^3}  \sum_{\alpha, \beta} A_{\alpha \beta} 
| h_{\nu \alpha j} v |^2   \ , 
\end{equation}
where the sum extends over the kinematically allowed decay channels, 
$v= \langle H \rangle$ = 174~GeV is the Higgs vev, $G_F$ is the Fermi
constant, and $A_{\alpha \beta}$ are $\order{1}$ coefficients that depend on the number of degrees of freedom associated to each mode.
Using the above equation, with
 at least one $h_{\nu \alpha j} \sim 10^{-7}$ we find that $m_j \gtrsim 1$~GeV, 
in order for $\nu_j$ to decay before BBN. 

Moreover, even if the decay of the $\nu_j$ occurs safely 
before BBN, it may lead to an increase of entropy density after the electroweak phase 
transition, which would dilute the baryon asymmetry. We have checked that   
this entropy increase is negligible for $m_j \gtrsim$ 10~GeV, when the $\nu_j$ decays
at $T \sim 500$~MeV, 
before the QCD phase transition. However it can be a concern for lower masses, 
$m_j \sim$ 1~GeV. In this case, the decay occurs after the QCD phase transition and the 
increase in entropy density can be up to order 10. This implies that the baryon asymmetry 
originally produced should be an order of magnitude larger, which could require 
Yukawa couplings close to the perturbative limit.   

\item[(ii)] Constraints from neutrino masses:

The seesaw contribution of $\nu_j$ to the light neutrino masses is  given by 
$(m_{L})_{ \alpha \beta} \sim h_{\nu \alpha j} h_{\nu \beta j} \frac{v^2}{m_j}$. 
Taking into account that 
$h_{\nu \alpha j} \gtrsim 2 \times 10^{-7}$, one has that $(m_L)_{\alpha \beta} [{\rm eV}] \gtrsim 1/m_{j} [{\rm GeV}]$. 
The strongest constraints on the absolute scale of neutrino masses are derived 
from cosmological observations, via their contribution to the energy density of the Universe 
and the growth of structure~\cite{planck13}. Since these bounds are very 
sensitive to the assumptions about the expansion history of the Universe and to the data included in the analysis, we choose the conservative 
upper bound on the sum of light neutrinos masses of roughly 1~eV, obtained by 
combining CMB and large scale structure  data when including several departures 
from the $\Lambda$CDM model~\cite{gonzalez10}.
This bound implies that $m_j \gtrsim 3$~GeV,  
unless there is a fine tuning among the phases of the Yukawa couplings of different species of $\nu's$, so that they give big contributions to $(m_L)_{\alpha \beta}$ with opposite signs that cancel each other. The atmospheric mass scale, 
0.05~eV, can be naturally obtained with $m_j \sim 20$~GeV, thus the two conditions
for our mechanism to work, 
$m_{j} \ll \tsph \sim 100$~GeV  and $h_{\nu \alpha j} \gtrsim 2 \times 10^{-7}$ are 
compatible with the observed light neutrino masses. Notice that these two requirements are
also needed when the helicity asymmetry in the singlet neutrinos is generated via 
neutrino oscillations~\cite{akhmedov98}.

Analogously, 
the heavy singlets $N_i$ also contribute to light neutrino masses an amount 
$h_{N \alpha i} h_{N \beta i} \frac{v^2}{M_i}$.
Barring accidental cancellations,  
$h_N \lesssim (10^{-5}-10^{-6}) \sqrt{M_i/1 \rm{TeV}}$
is consistent with present data, and also allows for the $N_i$ decay before BBN.

\end{itemize}

One may worry that the separation of the mass scales $m_j \ll M_i$ is not stable under 
radiative corrections, since there is not any symmetry protecting the small masses.
In fact, $\nu_j$ self-energy diagrams with virtual  
 $S_a$ and $N_i$ 
will induce Majorana masses for the $\nu_j$ at one loop of order 
\begin{equation}
m^{1-{\rm loop}} \sim
\frac{(\lambda_{aij})^2}{16 \pi^2}  M_i \log\left(\frac{M_i^2}{m_{Sa}^2}\right) \ .
\end{equation}
Thus for  $\lambda_{aij}$ of  $ {\cal O}(1)$, generically required for WB, 
we expect $m^{1-{\rm loop}} \sim 10^{-2} M_i$, which does not upset the condition
$m \ll 100$~GeV for $M_i$ of order few  TeV. Indeed, this loop contribution is 
naturally  of the correct size for helicitogenesis to work.


\section{Dynamics and evolution equations for the helicity asymmetry}
\label{sec:helicity}
In this section we will show how to calculate the helicity asymmetry in the $\nu_j$ sector and its partial transformation into a baryon asymmetry. For simplicity we will consider only one species of $N's$ and $\nu's$, $N\equiv N_1$ and $\nu \equiv \nu_1$, and hence we will omit the indices associated with the $\nu$ and $N$ sectors. As noted in~\cite{bento04}, in the  thermal bath rest frame isotropy implies that the spin density matrix is diagonal in the helicity basis. This allows to write a set of BEs for the populations of $\np{}$ and $\nm{}$ involving no coherences. Actually, the quantity of interest is the helicity asymmetry $Y_{\Delta \nu} \equiv Y_{\np{}}-Y_{\nm{}}$, where for any particle $X$ we define $Y_X \equiv n_X/s$ as the number density of $X$ normalized to the entropy density~\footnote{The population of $\nu$'s is in kinetic equilibrium due to different fast processes like scatterings with the $N$'s and Yukawa interactions with SM leptons. This allows for an easy integration of the momentum degrees of freedom, leading to simple Boltzmann equations for the number densities.}. 

The asymmetry $Y_{\Delta \nu}$ originates from interactions in the singlet sector of the model. In turn, this asymmetry is partially transferred to the lepton sector via the Yukawa interactions among $\nu$ and the SM lepton doublets. Finally the electroweak sphalerons transform part of the lepton asymmetry into a baryon one. A fairly good approximation is to consider that these different stages do not happen simultaneously: first $Y_{\Delta \nu}$ is generated while the DM annihilations freeze out and only then the Yukawa interactions and sphalerons act to get the final BAU. In other words, we neglect spectator processes during the generation of the helicity asymmetry. From the results of~\cite{nardi05} we expect that this type of approximation is accurate within factors not larger than $\sim 2$. We will also assume that thanks to the $N_i$-interactions described in the previous section, $Y_{N_i}$ follows an equilibrium distribution while the DM is annihilating.
Then $Y_{\Delta \nu}$ can be obtained from the following set of BEs (the details of the derivation of these BEs are very similar to those described in, e.g., the appendix~B of~\cite{bernal13}):
\begin{eqnarray}
 s z H(z) \frac{\dif Y_{\Delta \nu}}{\dif z} & = & \left(\ydyeqs{\chi} - 1 \right) 
 \epsilon \, \ghor{\chi \chi}{\nu N} 
  - \frac{Y_{\Delta \nu}}{Y_\nu^{eq}}  \Bigg[ 2 \ghor{\np{} N}{\nm{} N} \Bigg. \notag \\ \Bigg. & & + \ghor{\chi \chi}{\np{} N} - 4 \ghor{\np{} \np{}}{N N} - \ydyeq{\chi} \ghor{\chi \np{}}{\chi N} \Bigg] \; , \label{eq:beA}\\ 
s z H(z) \frac{\dif Y_{\chi}}{\dif z} & = & - 2 \left[ \ydyeqs{\chi} - 1 \right] \ghor{\chi \chi}{\nu N} \;. \label{eq:beB}
\end{eqnarray}
Here $z \equiv m_\chi/T$ and $H(z)$ is the Hubble rate. The reaction density  
$\ghor{a, b}{c, d}$
is the number of $\proname{a, b}{c, d}$ processes per unit time and volume, summing over all the degrees of freedom of the particles involved, including the helicity, unless this one is explicitly specified, as e.g. in $\ghor{\chi \chi}{\np{} N}$ which only involves $\nu$'s with positive helicity~\footnote{Note that the $N$'s are non-relativistic in the relevant epoch for baryogenesis, hence the chiral operators $P_R N$ and $P_L N$ in Eq.~\eqref{eq:lag} can create and destroy any helicity state of $N$. Therefore the helicity of the heavy sterile neutrinos does not play a major role and the rates are defined summing over the spin degree of freedom of $N$.}.
It is given by
\begin{equation*}
\ghor{a, b}{c, d}(z) = \frac{m_{\chi}^4}{64\,\pi^4\,z} \int_{x_{min}}^{\infty}dx \sqrt{x}\,\sigma_R(x\,m_\chi^2)\,K_1\left(z\,\sqrt{x}\right)\,,
\end{equation*}
where $x \equiv s/m_\chi^2$, $x_{min}={\rm Max}\left\lbrace\left(\frac{m_a+m_b}{m_\chi}\right)^2,\,\left(\frac{m_c+m_d}{m_\chi}\right)^2\right\rbrace$, and here $s$ is the center of mass energy squared~\footnote{We use the symbol $s$ both for the entropy density and for the center of mass energy squared. However it is always clear from the context which quantity we are referring to.}.
The reduced cross section $\sigma_R$ is related to the total cross section $\sigma$ via
\begin{equation*}
\sigma_R(s)=\frac{2\,\lambda(s,m_a^2,m_b^2)}{s}\sigma(s)\quad \text{and} \quad \lambda(s,m_a^2,m_b^2)\equiv \left(s-(m_a+m_b)^2\right)\left(s-(m_a-m_b)^2\right).
\end{equation*}

In the Eq.~\eqref{eq:beA} we have neglected the CP asymmetry in the decay of $S_{a}$, which is a good approximation if $m_{S_a} \gtrsim 2 m_\chi$ because that asymmetry would be washed out very efficiently. Hence the helicity asymmetry is generated mainly in the annihilation of DM and the CP asymmetry per annihilation appearing in Eq.~\eqref{eq:beA}, $\epsilon$, is defined as
\begin{equation}
\epsilon \equiv \frac{\Delta\gamma\left(\chi\chi\to\nu N \right)}{\ghor{\chi \chi}{\nu N}} \;,
\end{equation}
where
$\Delta\gamma\left(\chi\chi\to\nu N \right) = \ghor{\chi \chi}{\np{} N}-\ghor{\chi \chi}{\nm{} N}$
and $\ghor{\chi \chi}{\nu N}$  is the total annihilation rate, 
$\ghor{\chi \chi}{\nu N}=\ghor{\chi \chi}{\np{} N}+\ghor{\chi \chi}{\nm{} N}$.
As long as there are at least two species of scalars, $S_1$ and $S_2$, there is a contribution to the CP asymmetry at zeroth order in $m/m_{S_a}$ (where $m\equiv m_1$). Up to $\order{1}$ numerical factors, the reduced cross sections relevant
for the calculation of the CP asymmetry,  
  in the limit $m \to 0$, are given
 by \cite{bernal13}: 
 \begin{eqnarray}
& &\Delta\sigma_R\left(\chi\chi\to\nu N \right)= \frac{1}{8 \pi^2}\frac{\sqrt{s-4\,m_{\chi}^2}}{s^{3/2}\,(s-m_{S_1}^2)(s-m_{S_2}^2)}\times  \nonumber \\
& &\Bigg\lbrace 2 \lambda_{\chi 1}\lambda_{\chi 2}\,\text{Im}\left(\lambda_1\,\lambda_2^*\right) 
\left( |\lambda_1|^2 \frac{f_{S}(m_{S_1})+f_{V}(m_{S_1})}{s-m_{S_1}^2}-
|\lambda_2|^2 \frac{f_{S}(m_{S_2})+f_{V}(m_{S_2})}{s-m_{S_2}^2}\right)  \nonumber \\
&&-\text{Im}\left(\lambda_1^2\,\lambda_2^{*2}\right)\left( \lambda_{\chi 1}^2 \frac{f_{S}(m_{S_2})+f_{V}(m_{S_2})}{s-m_{S_1}^2}
-\lambda_{\chi 2}^2 \frac{f_{S}(m_{S_1})+f_{V}(m_{S_1})}{s-m_{S_2}^2} \right) \Bigg\rbrace\,,
\end{eqnarray}
and 
\be
\sigma_R\left(\chi\chi\to\nu N \right)= \frac{1}{8 \pi}s^{1/2}
\sqrt{s-4\,m_{\chi}^2} \, (s-M^2) 
\left\vert \frac{\lambda_{\chi 1} \,\lambda_1}{(s-m_{S_1}^2)}+
\frac{\lambda_{\chi 2} \,\lambda_2} {(s-m_{S_2}^2)} \right\vert^2\,,
\ee
where $\lambda_a \equiv \lambda_{a11}$. The loop functions, $f_S(m_{S_a})$ and $f_V(m_{S_a})$ (that also depend on $M \equiv M_1$), as well as the remaining density rates in Eq.~\eqref{eq:beA}, 
 can be found in appendix A of~\cite{bernal13}.

\begin{figure}[!t]
\centerline{\protect\hbox{
\epsfig{file=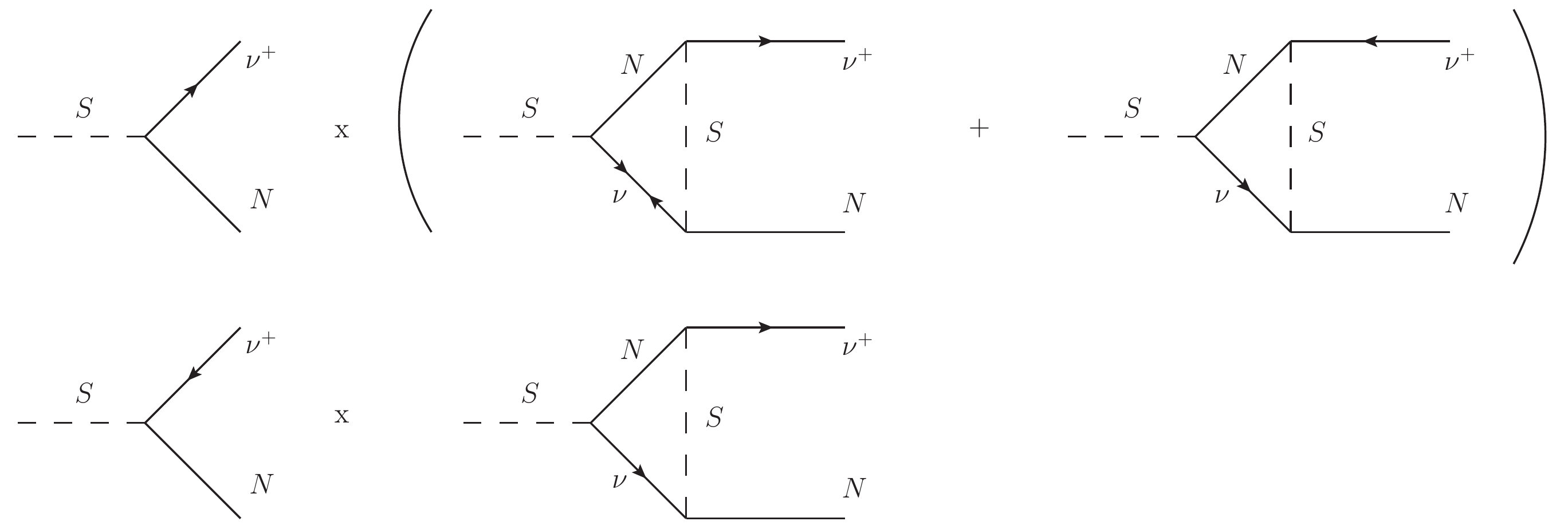,width=0.95\textwidth,angle=0}}}
\caption[]{Three of the lowest order contributions to the CP asymmetry in $S$-decays with only one species of scalars. Each one is obtained from the product between a tree level and a one loop diagram. The arrows indicate which of the two operators, $\lambda S \bar N P_R \nu$ (for ingoing arrows) or $\lambda^* S \bar \nu P_L N$ (for outgoing arrows), is used in each vertex. There are three more contributions obtained from the ones depicted by inverting the arrows in the loops. Each Majorana propagator ($\innerrightleftarrow$ or $\innerleftrightarrow$) as well as a ``wrong'' helicity emission ($\innerlongleftarrow \nu^+$) brings a suppression $m/m_S$ to the CP asymmetry.} 
\label{fig:1}
\end{figure}

In addition, the Majorana nature of $\nu$ and $N$ imply the existence of novel contributions to the CP asymmetry in annihilations as well as decays, with a CP odd phase given by $\miim{m M \lambda_a^2}$, see Fig.~\ref{fig:1}. Interestingly enough, this type of contributions requires just one species of real scalars, $S$, and therefore this brings a qualitative difference with typical baryogenesis scenarios. In particular, for the case of decays this means that the particle decaying and the one in the loop can be identical. However these contributions are suppressed by $\tfrac{m}{ 2 m_\chi}$ (for annihilations) or $\tfrac{m}{m_S}$ (for decays), which is a fairly small factor given that $m \ll \tsph = \, \order{100}$~GeV and $m_{\chi}, m_S \gtrsim 1$~TeV. Whether or not it is possible to have successful baryogenesis with this CP asymmetry is an interesting question, but it is beyond the scope of this work and it would probably require handling very large $\lambda_1$ couplings.

The second step in the approximation mentioned above is to analyze how the final helicity asymmetry obtained from the  BEs~(\ref{eq:beA}) and~(\ref{eq:beB}), $Y_{\Delta \nu}^f$, is transferred to the SM lepton sector. This occurs through different fast processes, all of them involving the Yukawa interactions among $\nu$ and the SM leptons. When $m=0$ it is possible to define a lepton number $L$ which is conserved by all these reactions, namely $L=L_{\rm SM} + L_\nu$, where $L_{SM}$ is the usual lepton number for SM fields, while $\np{}$ ($\nm{}$) is assigned $L_{\nu}=1 \, (-1)$. Hence the helicity can work as a lepton number for the sterile neutrinos $\nu$. Then the chemical equilibrium condition for the Yukawa interactions yields $\mu_\nu - \mu_\ell = \mu_H$, where $\mu_X$ is the chemical potential of the particle $X$. Taking into account also the whole set of relations among chemical potentials due to all the fast SM processes (including the electroweak sphalerons), the conservation laws, and the relation  among chemical potentials and density asymmetries~\cite{harvey90,nardi05}, one gets that $Y_B = a Y_{\Delta \nu}^f$. Here $a$ is a numerical factor whose value lies between $\sim 1/4$ and $\sim 1/3$ depending on how many independent fast Yukawa interactions there are among the $\nu_j$ and $\ell_\alpha$. 

When $m \neq 0$, $L$ is not conserved. However if $m \ll T$ the rate of processes violating $L$ is suppressed by $(m/T)^2$ with respect to the rate of reactions conserving L. Hence a net lepton asymmetry can be transferred to the SM sector. As an example consider the top-quark scattering $\proname{\nu \ell_\alpha}{Q_3 \bar t}$, with $t$ the right top and $Q_3$ the third generation SU(2) quark doublet. If $\nu$ has negative helicity the process conserves $L$, i.e. it allows to transfer helicity asymmetry into SM lepton asymmetry with the ``correct'' sign. Instead, if $\nu$ has positive helicity, the process violates $L$ and leads to a washout of $L_{\rm SM}$ and $L_\nu$. The amplitudes for these reactions are equal to
\begin{eqnarray}
\abss{\mathcal{M}(\proname{\ell_\alpha \nm{}}{Q_3 \bar t})} &=& \frac{\abss{h_{\nu \alpha}} \abss{h_t} \, s}{(s-m_H^2)^2} \, (p^0 + \abs{p})(k^0+\abs{k}) \sin^2 \frac{\theta}{2} \; ,\label{eq:Mnum}\\
 \abss{\mathcal{M}(\proname{\ell_\alpha \np{}}{Q_3 \bar t})} &=& \frac{\abss{h_{\nu \alpha}} \abss{h_t} \, s}{(s-m_H^2)^2}  \, (p^0 + \abs{p})(k^0 + \abs{k}) \cos^2 \frac{\theta}{2} \left(\frac{m}{p^0 + \abs{p}}\right)^2 \; , \label{eq:Mnup}
\end{eqnarray}
where $p^\mu$ ($k^\mu$) is the momentum of $\nu$ ($\ell_\alpha$) and $\theta$ is the angle between $\stackrel{\to}{p}$ and $\stackrel{\to}{k}$. The reaction densities are obtained from the thermal average of the cross sections in the thermal bath rest frame. The integration can be done analytically only for  $\proname{\ell_\alpha \nm{}}{Q_3 \bar t}$,
\begin{equation*}
\ghor{\ell_\alpha \nm{}}{Q_3 \bar t} = \frac{1}{4} \frac{1}{(2 \pi)^5} \abss{h_{\nu \alpha}} \abss{h_t} T^4 \; ,
\end{equation*}
where $h_{\nu \alpha}$ and $h_t$ are the Yukawa couplings of $\nu$ and $t$, respectively, and we have neglected the mass of the Higgs in the propagator. Nevertheless, the important point is that due to the factor $\big(\tfrac{m}{p^0 + \abs{p}}\big)^2$ in Eq.~\eqref{eq:Mnup}, $\tfrac{\ghor{\ell_\alpha \np{}}{Q_3 \bar t}}{\ghor{\ell_\alpha \nm{}}{Q_3 \bar t}} = \order{(m/T)^2}$, hence the washouts induced by $\proname{\ell_\alpha \np{}}{Q_3 \bar t}$ will be negligible as long as $m \ll \tsph$, as stated before. 

As an illustration of the allowed parameter space we provide in table~\ref{tab:bp} a benchmark point that has been obtained solving the BEs~\eqref{eq:beA} and~\eqref{eq:beB}. At the beginning of this section we mentioned that a more accurate calculation of the final baryon asymmetry should take into account that the generation of helicity asymmetry and its partial conversion into net SM lepton and baryon numbers occur simultaneously. Such degree of accuracy is outside the scope of this work, but we note that a convenient way to take into account these spectator effects is to set a BE for the quantity $B-L_{SM}-L_\nu$. 
\begin{table}[!t]
\centering
\begin{tabular}{ l|cccc|cccc l }
\toprule
Parameter & $m_{\chi}$ & $m_{N}$&  $m_{S_1}$& $m_{S_2}$  & $\lambda_{\chi 1}$ & $\lambda_{\chi 2}$ & $\lambda_{1}$ & $\lambda_{2}$\\
\midrule
Benchmark & 2 & 3  & 6 & 6.6  & 1.5 & 1.65 & 1.7 & 1.8 \\
\bottomrule
\end{tabular}
\caption{Benchmark point for the WB model. The masses are in TeV and the phases of the Yukawa couplings have been chosen to maximize the CP asymmetry $\epsilon$. The remaining couplings and masses must be in the ranges specified in section~\ref{sec:model}.} \label{tab:bp}
\end{table}


\section{Other realizations of helicitogenesis}
\label{sec:other}

\subsection{WIMPy leptogenesis with spontaneous $U(1)_L$ symmetry breaking}

A drawback of the minimal model we have described is that there is no justification for the necessary hierarchical spectrum of sterile neutrino masses and couplings. 
Actually those hierarchies would suit very well if  the singlet fields were charged under a conserved lepton number. The assignment $L=0,1,1$, and 1/2 to $N_i, \nu_{R \,j}, S_a$, and $\chi$, respectively, would imply that $m_j=\lambda_{\nu a i j}=\lambda_{N a i j}= h_{N \alpha i}=0$, which perfectly fulfills the requirements of the helicitogenesis mechanism~\footnote{In this case the $N_i$ can decay before BBN thanks to the large couplings $\lambda_{a i j}$ and the $S_a - H$ mixing after electroweak and $U(1)_L$ symmetry breaking.}.  
The most general $L$-conserving Lagrangian can be written as 
\begin{equation}
\label{eq:lag2}
\begin{split}
-L & = - L_{SM} - L_{kin} + V(S_a,H) 
+  m_\chi \overline\chi \chi +  \frac{1}{2} M_i \overline N_i N_i  \\ 
&  + \frac{1}{2}  \Big\{ \lambda_{\chi a R} S_a^\dag  \overline{\chi^c} P_R \chi 
 + \lambda_{\chi a L} S_a^\dag \overline{\chi^c} P_L \chi 
+ {\rm h.c.}   \Big\} + \Big\{   
 \lambda_{a i j} S_a^\dag \overline N_i P_R \nu_j  
  + h_{\nu \alpha j} \tilde H \overline \ell_\alpha P_R \nu_j + {\rm h.c.}\Big\} \; ,
\end{split}
\end{equation}
with 
\begin{equation}
\label{eq:V}
V(S_a,H) = m^2_{S a} S_a^\dagger S_a + 
\lambda_{Hab} (H^\dagger H )  (S_a^\dagger S_b)
+  \lambda_{abcd} (S_a^\dagger S_b) (S_c^\dagger S_d) +  {\rm h.c.}
\end{equation}

The DM field $\chi$ is now a Dirac fermion, and $\chi^c = C \overline{\chi}^T$. 
Recall that WB requires at least two scalar fields $S_a$.
After the complex scalars acquire a non zero vev, 
lepton number gets spontaneously broken and $\chi$ splits into two Majorana fermions 
$\chi_1, \chi_2$ with masses
\begin{equation}
m_{\chi_1,\chi_2} = 
\frac 1 2 \left\{ \mu_L + \mu_R \pm
\sqrt{   (\mu_L -\mu_R)^2 + 4 m_\chi^2 }  \right\} \ ,
 \end{equation}
where $\mu_{L,R} \equiv \sum_a \lambda_{\chi a L,\chi a R} \, u_a$ 
and $u_a = \langle S_a \rangle$. Notice that in this model we do not need an additional $Z_2$ symmetry, as it is usually the case to avoid the DM decay: the lightest $\chi_i$ is stable because 
of a $Z_2$ symmetry which is an unbroken remnant of the global $U(1)_L$, as in~\cite{thomas11}.
 
The light neutrino masses would be obtained via a double seesaw~\cite{mohapatra86} 
mechanism;
 once the electroweak symmetry is also broken, the SM doublet neutrinos $\nu_\alpha$ 
and the sterile ones,  $\nu_j, N_i$,  mix and 
the mass matrix in the ($\nu_{\alpha} , \nu_j, N_i$)  basis becomes:
\begin{equation}
\cal {M} = \left(
\begin{array}{ccc}
0 & h_\nu v & 0 \\
h^T_\nu  v & 0 &   \lambda^T_a u_a  \\
0 &   \lambda_a u_a  & M
\end{array}
\right) \ , 
\end{equation}
where $v= \langle H \rangle$,   
the matrix elements of $\lambda_a$ ($h_\nu$) 
are the Yukawa couplings $\lambda_{aij}$ ($h_{\nu \alpha j}$), 
and a sum over repeated indices 
is understood.
 In the limit $\lambda_a u_a \ll M$, the singlets 
$\nu_j$ acquire a mass given by 
\begin{equation}
\label{eq:mj}
m =  (\lambda^T_a u_a) M^{-1}  (\lambda_a u_a)   \ll M \ , 
\end{equation}
while the mass matrix  of the three light neutrinos is  
\begin{equation}
\label{eq:mL} 
m_{L} =  h \, m^{-1} h^T  v^2 \ .
\end{equation}
Therefore the smallness of the $\nu_j$ masses is due to a seesaw mechanism involving just the 
SM singlet leptons.

There are different variants of this scenario, depending on whether lepton number is a global or local symmetry, and the time of spontaneous breaking, denoted by the temperature $T_L$. 
We first consider the case of global $U(1)_L$. If lepton number is broken 
after the DM freeze out, WB   
does not work because the $\chi$ field, being charged,  can also hold an asymmetry. In turn this asymmetry induces a washout proportional to $\tfrac{\ghor{\chi \chi}{N \nu}}{n_\chi^{eq} H(z)}$, which freezes out at the same moment as the annihilation of DM, violating one of the basic requirements of WB~\cite{cui11}~\footnote{The baryon asymmetry is roughly given $Y_B \sim \tfrac{\epsilon}{2} [Y_\chi(z_w)-Y_\chi(\infty)]$, where $z_w$ is the value of $z$ at which washouts freeze out and $Y_\chi(\infty)$ is the relic DM density normalized to the entropy density. Given that $\Omega_{DM} \sim 5 \,\Omega_{B}$ and $m_\chi \gtrsim 1$~TeV, it is clear that all washout processes must freeze out before DM annihilations, when $Y_{\chi}$ is several orders of magnitude above its final value $Y_\chi(\infty)$.}.

However, it may be  possible that WB occurs via 
helicitogenesis when $U(1)_L$ is already broken, i.e.,  
$T_L > T_h > T_{sfo}$, being 
$T_h$ the temperature at which the $\nu_j$ helicity asymmetry
is generated. 
After $U(1)_L$ breaking, 
 the mass of the singlet scalars $S_a$ is  $m_{Sa} \sim \lambda u_a$, with 
 $\lambda$ the quartic coupling in the scalar potential. Thus, for $T_L \sim$  few 
 TeV  and   some quartic couplings $\lambda$ of $\order{1}$, 
 $m_{Sa} >  m_{\chi_1} \gtrsim $ 1~TeV can be easily achieved.  
The sterile neutrinos $\nu_j$ acquire a mass given by Eq.~(\ref{eq:mj}), 
therefore the condition $m_j \ll \tsph \sim$ 100~GeV 
needed to generate the helicity asymmetry in $\nu_j$, 
leads to $(\lambda_a u_a)^2 /M \ll 100$~GeV.
 For instance,  assuming  $u_a \sim M=$ 1~TeV,   
$\lambda_a$ of order   0.2  is required to obtain $m_j \sim $ 40~GeV.
On the other hand,  the Yukawa couplings 
$\lambda_{aij}$
should be sizeable, of ${\cal O}$(1), to have enough CP violation and get the correct DM relic abundance, so there is some tension between these two requirements for WB.
Since  the neutrino masses depend only on  the couplings $\lambda_a$
 while  a combination of 
 $\lambda_a$ and $\lambda_a^*$ appears 
 in the CP-asymmetry, it is conceivable  that some cancellations due to phases 
 allow to satisfy all constraints in certain regions of the parameter space
 with only two singlet scalars.
 Alternatively, in the presence of three scalars  
it may happen that 
$u_1, u_2  \ll u_3$ but  $\lambda_{1}, \lambda_{2}  \gg \lambda_{3ij}$, achieving a large enough CP asymmetry through the couplings $\lambda_{1,2}$  and getting  $m_j \ll \tsph$ 
without  accidental cancellations. 
Moreover,  the vev's at $T_h > T_{sfo}$ may  be different from 
the vev's at $T=0$, as in the singlet Majoron model~\cite{cline09}, helping 
 to enlarge the allowed parameter space. 
We thus conclude that WB seems feasible in 
the present framework  if
$T_L > T_h > T_{sfo}$.

The spontaneous breaking of a global symmetry leads to a massless Goldstone boson, 
the Majoron,
\begin{equation} 
J = \sum_a {\frac{u_a}{u}} \,   {\rm Im} \,S_a  \ , 
\end{equation} 
with $u = \sqrt{\sum_a u_a^2}$.
However, non-perturbative gravitational effects are expected to explicitly break global symmetries and provide a mass to the Majoron~\cite{akhmedov92}. 
If this mass is  $m_J  \lesssim$ few hundred GeV, 
 processes mediated by $N$ such as $\nu_i^+  J  \rightarrow \nu_j^- J $ could lead to 
a fast washout of the $\nu$ helicity asymmetry and a 
more detailed analysis is required.

This potential problem can be avoided by promoting lepton number to a gauge symmetry. 
In this framework,
only the case $U(1)_{B-L}$ 
is anomaly free without requiring new exotic fermions to cancel anomalies. 
Then,  there are additional constraints due to the searches of the 
extra $Z'$ gauge boson at LEP, Tevatron and LHC. 
While LHC searches for heavy resonances depend both on the $U(1)_{B-L}$ coupling 
strength $g_{B-L}$ and $Z'$ mass~\cite{cms10, ATLAS-CONF-2013-017}, 
limits from LEP II imply a model independent bound on the  vev
$u = M_{Z'}/(2\, g_{B-L}) \gtrsim 3$~TeV~\cite{Carena:2004xs}~\footnote{Note that if the $Z'$ can decay into sterile neutrinos, the LHC limits may be relaxed~\cite{abdelalim14}.}.
Thus in the gauged case $U(1)_{B-L}$ is necessarily broken 
before the electroweak phase transition, and the results discussed above 
for $T_L > T_h > T_{sfo}$ apply.  
Now there is no Majoron, and the only concern 
would be that the cross sections of the new lepton-number-conserving annihilation
channels
induced by the  $Z'$ boson are not too large, 
so a significant fraction of DM annihilations still proceed 
through lepton-number-violating processes with $\nu$ in the final state, leading to 
a sizeable helicity asymmetry in the $\nu$ population. 
This requirement is easy to satisfy, since $g_{B-L}$ as well as $M_{Z'}$  are free parameters.

\subsection{Helicitogenesis in WIMP decay}

It is also possible to realize baryogenesis via helicitogenesis 
if the matter-antimatter asymmetry does not originate in the DM annihilations, but in 
the out-of-equilibrium decay of the heavy SM singlets, namely 
 the scalars 
$S_a \rightarrow N_i \nu_j$, if $m_{S a} > M_i$, or the heavy fermions
$N_i \rightarrow S_a \nu_j$, if $m_{S a} < M_i$. 
In this case, the conditions $m_{Sa} >  m_{\chi} $ and 
$m_\chi \lesssim M_i < 2 m_\chi$ are 
 not required.
Although the connection between the DM and baryon abundances is 
in principle lost,
it provides a new mechanism to produce the desired helicity asymmetry in the light
singlets, $\nu$, so we also discuss this scenario.
In the following, we assume that either all the singlet fermions 
are lighter than the lightest scalar, which we denote by $S_1$, or both scalars 
are lighter than the lightest heavy fermion, denoted by $N_1$. 
These two illustrative scenarios contain all relevant physical features, so 
more involved  mass spectra  will not introduce any significant new ingredient in our analysis.

We first consider the possibility that the 
$\nu_j$ helicity asymmetry is generated after $U(1)_L$ breaking, at temperatures 
$T_L > T_h \sim M_d/10 > \tsph$, with $M_d$ the mass of the decaying particle, 
$N_1$ or $S_1$. 
The following discussion applies to both, global and gauged lepton number, although 
in the first case if the Majoron is too light it could wash out the helicity asymmetry.

The usual requirements for standard leptogenesis at low temperatures should be satisfied, 
namely
$M_d$  $\gtrsim$ 1~TeV  and tiny Yukawa couplings of the decaying particle, typically of order 
$\lesssim 10^{-7}$.
This last requirement has different implications depending on which 
particle generates the helicity asymmetry, $S_1$ or $N_1$. 
In the first case, since the Yukawa couplings $\lambda_{1ij}$ are tiny, the mass of all the singlets 
$\nu_j$ is mainly generated by the vev and couplings of the scalar 
$S_2$ in Eq.~(\ref{eq:mj}).
In the second one, $\lambda_{a1j} \lesssim 10^{-7}$ for all $a,j$ leads to a negligible contribution 
of $N_1$ to the $\nu_j$ masses, so in order 
to get at least two massive $\nu$'s able to  generate the observed SM neutrino masses, two more $N$'s with sizeable Yukawa couplings (of order $0.1$) are needed.
A big enough CP-asymmetry is obtained for couplings $\gtrsim 10^{-3}$, which is a weaker 
constrain. If there are only three heavy $N_i$, 
one of the $\nu_j$ remains very light,
with $m_j \lesssim 10^{-11} (u/{\rm TeV})^2$~GeV, 
 so it should have  Yukawa couplings to the SM particles $h_{\nu\alpha j} \ll 10^{-7}$
to avoid too large contributions to the SM neutrino masses. 
This is not a problem, provided some of the other $\nu$ Yukawa couplings
are $\gtrsim 2 \times 10^{-7}$, to efficiently transfer the helicity asymmetry to a
lepton number asymmetry in the SM doublets $\ell_\alpha$. 
If there are more than three $N$'s, all the $\nu$ masses can be 
$\gtrsim$ 10~GeV.

Using the results of \cite{racker13}, we conclude that 
 successful baryogenesis is realized for
$m_{S1} \sim $ few TeV, $M_i \sim $ (0.5 - 1)~TeV and Yukawa couplings 
$\lambda_{1ij} \lesssim 10^{-7}$, $\lambda_{2ij} \sim$ 0.1.  
Such values also lead naturally to $m_j \sim $ 10~GeV  
for $u_a \sim M_i$. 
If $m_{S1} < M_i$, leptogenesis could occur in the decay $N_1 \rightarrow S_a \nu_j$, 
for a similar range of masses and couplings of the particles involved, just exchanging
the roles of $S_a$ and $N_i$. 
This situation seems more contrived, because a largish $M_i$ tends to give too small $\nu_j$ masses   
from Eq.~(\ref{eq:mj}), which in turn will produce a light neutrino mass above the cosmological 
upper limit $\sim$ 0.3~eV~\cite{planck13}.
In summary, helicitogenesis via WIMP decay  seems likely to occur in 
a large region of the parameter space if $T_L > T_h > T_{sfo}$.
A more exhaustive analysis is beyond the scope of this work.

When $U(1)_L$ breaking takes place at temperatures $T_L < T_h \sim M_d/10$,
baryogenesis occurs in a $B-L$ conserved fashion.
Moreover, the $\nu_j$ are exactly massless  at $T_h$, 
so the requirement $m_j \ll T_{sfo}$ can be relaxed and 
their helicity asymmetry is in fact a lepton number asymmetry.
Generically, we expect $m_{S a} \sim T_L $, thus it seems  more natural 
to consider the decay process  $N_1 \rightarrow S_a \nu_j$. 
However, since the scalars $S_a$ are charged under $U(1)_L$, 
an equal and opposite lepton number asymmetry is generated in the scalar sector 
between  $S_a$ and $S_a^*$. Therefore this case is not an example of the low scale baryogenesis mechanism 
proposed here, namely massive decay or annihilation products which do not store asymmetry,
such as real scalars or heavy Majorana fermions.
In fact, within this model the suppression of washouts due to massive decay products is not 
effective. There are other possibilities of getting 
 successful leptogenesis,  like very heavy neutral leptons, 
$M_i \gtrsim$ 100~TeV, an initial  thermal abundance of $N_1$ followed by a late decay~\footnote{An initial thermal abundance of $N_1$ can be produced if one of the scalars $S_a$ is heavier 
than $N_1$ and has significant Yukawa couplings $\lambda_{a1j}$, while the late decay occurs if the 
Yukawa couplings of $N_1$ to the lightest scalar $S_1$ are tiny, $\lambda_{11j} \ll 10^{-7}$.},
 or two almost degenerate $N_i$~\cite{piu}.
 In general, the  neutrino masses $m_j$ tend to be too small,  unless the scalar vevs are unusually 
 large, $u_a \sim M_i$, 
 so we do not discuss further this possibility.


\section{Conclusions and outlook}
\label{sec:conclusions}

We have shown how baryogenesis can be achieved from the annihilation or decay of heavy particles - masses ${\cal O}(1)$~TeV - into sterile neutrinos. One of these neutrinos, $N$, must be also ``heavy'' ($M \gtrsim 0.5$~TeV), providing a Boltzmann suppression of washouts which can be very fast at these low scales for thermal baryogenesis. The other neutrino, $\nu$, must be relatively ``light'' ($m \ll 100$~GeV), so that its mass  is  negligible during the annihilation or decay epoch, allowing for an helicity asymmetry to be generated. This asymmetry is partially transferred to SM leptons via Yukawa interactions and subsequently reprocessed into a baryon asymmetry via the electroweak sphalerons.

Given that one of the sterile neutrinos must have $m < $ few x 10~GeV and some of the Yukawa couplings must be larger than $h_\nu \gtrsim 2 \times10^{-7}$, the mass of at least one of the light SM neutrinos must be larger than few x 0.01~eV -barring phase cancellations in the mass matrix-, which is intriguingly close to the mass scales set by neutrino oscillations.

 We have studied a realization of the mechanism in the framework of WB, where there is a relation between the BAU and DM. Since the sterile neutrinos responsible for neutrino masses play fundamental roles in the generation of the BAU and the freeze out of DM, helicitogenesis from the annihilation of DM yields a connection between neutrino masses, the BAU and DM.

The required pattern of sterile neutrino masses appears naturally in the so-called double seesaw 
mechanism, where the smallness of the $\nu$ masses can be due to a $U(1)_L$ symmetry
spontaneously broken. 
Thus, we have constructed an extended $U(1)_L$ symmetric 
double-seesaw model, including also fermionic DM and two SM singlet scalars, all of them  charged under lepton number.
We have shown that it is possible to reconcile the helicitogenesis requirements with the measured light neutrino masses provided that $U(1)_L$ breaks spontaneously prior to 
DM freeze out, or heavy particle decay, and hence before the electroweak phase transition. 
Within this framework,  it seems feasible to have 
successful WB and explain the observed light neutrino masses with  
 DM and sterile neutrino couplings to the singlet scalars close to ${\cal O}(1)$. 
The suppression of fast washouts is also at work when 
the helicity asymmetry in the ``light'' sterile neutrinos $\nu$ 
is generated  during the out-of-equilibrium decay of the heavy states, 
namely the singlet fermions $N$ or the singlet scalars. 
 In this case, 
 the decaying particle should have tiny Yukawa couplings, to achieve the out-of-equilibrium condition at low temperatures, $T \sim \order{1}$~TeV, and 
the direct connection between the baryon asymmetry and  the DM relic abundance is lost.
 The presence of a light Majoron associated to the breaking of the 
global $U(1)_L$ symmetry is a potential problem, 
which can be solved by gauging $U(1)_{B-L}$.

The mechanism that we propose has certain similarity with baryogenesis via neutrino oscillations, in that the source of the baryonic asymmetry is an helicity asymmetry in the sterile neutrino sector. 
As a consequence, the three requirements $m \ll 100$~GeV,  fast Yukawa interactions 
and  generation of the helicity asymmetry before sphaleron freeze out apply in all cases,
because the first one allows for the existence of the helicity asymmetry itself and the 
last ones take care of the efficient transfer of the asymmetry to the baryonic sector. 
However the generation of the $\nu$ helicity asymmetry described here is a completely different 
process, which involves the CP-violating  annihilation or decay of new WIMPs 
to the sterile neutrinos $\nu$.

Notice that the suppression of the washout processes due to a heavy decay or annihilation product which does not store asymmetry is very general, and not only applies to the scenarios we have described here. E.g., it will also occur if the other decay product is a SM lepton, in which case the lepton asymmetry is generated directly in the out-of-equilibrium decay. This possibility can be realized generalizing the inert doublet model~\cite{ma06} with  two inert scalars coupled to SM leptons and  sterile neutrinos. Leptogenesis would occur in the decay of the lightest scalar at the TeV scale, while the second inert scalar is necessary to have CP violation. Alternatively, the role of the massive decay or annihilation product could also be played by a real scalar.

From the phenomenological point of view, the models that we have discussed 
involve new particles at the 
TeV scale, so in principle they can be tested in current or near-future experiments.
 It would be worth to analyze whether 
 the prospects for detecting the sterile neutrinos, generically very difficult in 
Type I seesaw models, 
are improved by their additional interactions 
with the singlet scalars, which in general mix with 
the SM Higgs field, or with the $Z'$ boson in the $U(1)_{B-L}$ gauged case.
With respect to DM detection, the phenomenology of WB models has been 
extensively analyzed neglecting the mixing among the 
SM Higgs and the extra scalar singlets~\cite{cui11,bernal12}. 
The observable signatures are not very promising when DM annihilates to leptons, 
 however the impact of the mixing within the scalar sector and of the  
 $Z'$ interaction deserves further investigation.

\section*{Acknowledgments}
We thank Juan Herrero Garc\'\i a, Arcadi Santamar\'\i a and Aaron Vincent for illuminating discussions. 

This work has been supported by the Spanish MINECO Subprogramme Juan de la Cierva and it has also been partially supported by the Spanish MINECO grants FPA2011-29678-C02-01 and Consolider-Ingenio CUP (CSD2008-00037), and by Generalitat Valenciana grant PROMETEO/2009/116. In addition we acknowledge partial support from the  European Union FP7  ITN INVISIBLES (Marie Curie Actions, PITN- GA-2011- 289442).
\bibliographystyle{JHEP}
\bibliography{referencias_leptogenesis2}

\end{document}